\documentclass[3p,lefttitle,sort&compress,fleqn]{elsarticle}
\makeatletter
\def\ps@pprintTitle{%
  \let\@oddhead\@empty
  \let\@evenhead\@empty
  \def\@oddfoot{\reset@font\hfil\thepage\hfil}
  \let\@evenfoot\@oddfoot
}
\makeatother

\usepackage{amsmath}
\usepackage{booktabs}
\usepackage{multirow}
\usepackage{dcolumn}
\newcolumntype{d}{D{.}{.}{-1}}

\usepackage{amssymb}
\usepackage{amsthm}

\theoremstyle{definition}

\usepackage{hyperref}
\hypersetup{colorlinks}

\DeclareGraphicsExtensions{.eps}

\begin{document}

\begin{frontmatter}

\title{Resonant excitation of the bushes of nonlinear vibrational modes in monoatomic chains}
\author{George Chechin}
\ead{gchechin@gmail.com}
\author{Galina Bezuglova}
\address{Institute of Physics, Southern Federal University, 194 Stachki ave., Rostov-on-Don 344090, Russia}

%\date{\today}

\begin{abstract}

Intermode interactions in one-dimensional
nonlinear periodic structures have been
studied by many authors, starting with the
classic work by Fermi, Pasta, and Ulam (FPU)
in the middle of the last century. However,
symmetry selection rules for the energy
transfer between nonlinear vibrational modes of different symmetry, which lead to the
possibility of excitation of some \emph{bushes} of such
modes, were not revealed. Each bush
determines an exact solution of nonlinear
dynamical equations of the considered
system. The collection of modes of a given
bush does not change in time, while there is a
continuous energy exchange between these
modes. Bushes of nonlinear normal modes
(NNMs) are constructed with the aid of group-
theoretical methods and therefore they can
exist for cases with large-amplitude atomic
vibrations and for any type of interatomic
interactions. In most publications, bushes of
NNMs or similar dynamical objects in one-
dimensional systems are investigated under
periodic boundary conditions. In this paper,
we present a study of bushes of
NNMs in monoatomic chains for the case of
fixed boundary conditions, which sheds light
on a series of new properties of intermode interactions in such systems, as well as a method for constructing
bushes of NNMs by continuation of
conventional normal modes to the case of
large atomic oscillations. The
present study was carried out for the chains with the
Lennard-Jones interatomic potential, but methods developed in this paper are valid for
arbitrary monoatomic chains.

\end{abstract}


\end{frontmatter}

\section{\label{sec1}Introduction}

The study of intermode interactions in one-dimensional chains has
a long history, starting with the famous work by Fermi, Pasta and
Ulam \cite{FPU1955}. In this work, a mathematical model was
introduced, which represents a chain of identical weights
(hereafter referred to as particles or atoms) connected by
identical springs. Stretching of such spring generates an elastic
force, which is described by Hooke's law with nonlinear additions
of the second (FPU-alpha model) or third (FPU-beta model) order.

This simple mechanical model had a huge
impact on the modern nonlinear dynamics and
led to the discovery of such dynamical objects
as solitons \cite{ZabuskyKruskal1965, Zabusky1981} and such phenomenon as
dynamical chaos \cite{Chaos2005}.

The present work is devoted to dynamical objects, which are called
\emph{bushes} of nonlinear normal modes (NNMs). Speaking about the
general theory of bushes of NNMs (hereafter bush theory), which
can be apply to study of atomic vibrations of large amplitudes in
any systems with discrete symmetry one must keep in mind the
following properties of these dynamical objects.

Since their existence is due only to the symmetry properties, they
can be found in any complex structures with any type of
interatomic interactions.

Let us recall some works that preceded the appearance of the
general bush theory. For the FPU-beta model, some sets of modes
that persist during the time evolution of the system were found in
Ref.~\cite{Ruffo1997}. After this paper by Poggi and Ruffo,
several papers of other authors were published in which intermode
interactions were studied based on the specifics of interatomic
interactions in the FPU chains \cite{art23}, \cite{art24},
\cite{art25}, \cite{art26}. Only in Refs. \cite{art27},
\cite{art28} Rink used some group-theoretical methods for
constructing   exact dynamical objects consisting of several
vibrational modes, however, this work was performed again
precisely for one-dimensional FPU-chains.

In our research group, the general theory of
bushes of NNMs was developed from
essentially different positions. We started study
with our theory of complete condensate of
order parameters for structural phase
transitions \cite{pss1989} and generalized it to the case of
nonlinear dynamics of any systems with
discrete symmetry \cite{art10}.

In Ref.~\cite{art9}, the selection rules were found
for excitation transfer between vibrational
modes of different symmetry. It is these
selection rules that underlie the geometric
aspect of the bushes of NNMs. A detailed
description of the bush theory is given in Ref.~\cite{art10}.
In a series of subsequent works, we studied
bushes of NNMs in various systems with
different groups of point and space symmetry
(see references to these works in the Ref.
\cite{ChR2022}).

However, despite the large number of works of such type, the
methods of \emph{excitation} of bushes of NNMs both in the
framework of computational or real physical experiments were not
discussed. It is this problem that the present work is concerned
with.

Bushes of NNMs are \emph{exact solutions} of nonlinear dynamical
equations lying on invariant manifolds defined by the symmetry of
the considered system. These solutions cannot be obtained within
the framework of any perturbation theory.

Each bush defines an invariant manifold, on
which lies some exact solution of the nonlinear dynamical equations
of motion of the considered system.

There are no general methods for finding invariant manifolds. But
if the given system has some discrete symmetry, i.e. it is
described by any point or space group, such a method is offered by
the bush theory of NNMs. It was shown in \cite{art10} that the
subspace of the phase space, which is invariant with respect to
the subgroup $G_j$  of the system`s symmetry group $G_0$, is simultaneously
an invariant manifold of the considered nonlinear dynamical
system. It can be found by solving linear algebraic
equations of the form:

\begin{equation}\label{eqzv}
\hat{G} * X = X,
\end{equation}

where the multidimensional vector $X$ determines the full set of
displacements of all atoms of the system from their equilibrium
positions\footnote{The symmetry elements of the group $G_0$ act on
the vectors of the three-dimensional Euclidean space, while
$\hat{G}$ is the group of induced operators that act on the
multidimensional vectors $X$ of atomic displacements.}.

Thus, the problem of finding invariant manifolds of a nonlinear
dynamical system is reduced to a purely geometric problem of
finding subspaces of the phase space that are invariant with
respect to subgroups of the symmetry group of the considered
system, and the problem of finding all possible bushes is reduced
to solving systems of the form (\ref{eqzv}) for all subgroups
$G_j$ of the group $G_0$.

Bushes of NNMs can be used for describing large amplitude atomic
vibrations in systems with complex structures corresponding to
different point and space symmetry groups.

In Refs. \cite{art14}, \cite{art15}, \cite{art17}, \cite{art18}, \cite{art34}, \cite{art51}], bushes of vibrational modes in
various two-dimensional and three-dimensional systems were found and
properties of these dynamical objects were
investigated. Bushes of nonlinear modes in
one-dimensional carbon chains (carbynes)
were studied in \cite{art16}, \cite{art102}, \cite{art103}, \cite{art104}.

For the present work we need to use some results of the paper
\cite{ChR2022} devoted to studying bushes of NNMs in arbitrary
monoatomic chains with \emph{fixed}  ends. In that paper, the
method for constructing bushes of vibrational NNMs from the
conventional (linear) normal modes was developed.

As is well known, normal modes are obtained for the case of small
vibrations of mechanical systems in the framework of harmonic
approximation. Each normal mode has the form of a product of the
\emph{atomic displacement pattern} [$a_1$, $a_2$, $a_3$,... $a_n$]
and some sinusoidal function of time $f(t)$:

\begin{equation}\label{eq1}
[a_1, a_2, a_3, \ldots, a_n]\,f(t),
\end{equation}

\noindent
where the constant coefficients $a_i$ determine
displacements of all atoms of the considered
system from their equilibrium positions.

Speaking about the excitation of a given normal mode, we mean the
following mathematical procedure: we set sufficiently small values
of the atomic displacements (to validate the harmonic
approximation) in accordance with formula (\ref{eq1}) and set all
the initial atomic velocities to be zero. Then we solve the Cauchy
problem for the system of nonlinear equations of motion by some
numerical method on a certain finite time interval. The presence
of anharmonic terms in the power expansion of the potential energy
leads to the appearance of a weak interaction between normal modes
and this is the reason for the appearance of a certain bush of
normal modes.

The above procedure for exciting vibrational bushes is quite
simple to carry out in  \emph{computational} experiments, but it
is impossible to carry out in  physical experiments.

The purpose of the present work is to develop a mathematical
scheme for exciting vibrational bushes based on such resonant
actions on the considered system that, in principle, could be
carried out within the framework of  physical experiments.

Further presentation of the paper is organized as follows. In
Sec.~\ref{sec2} we present two mathematical models describing the
resonant excitation of the oscillatory bushes in monoatomic
chains. Section~\ref{sec3} is devoted to the excitation of
vibrational modes in the chain with two mobile atoms. In
Section~\ref{sec4} we investigate the excitation of bushes of NNMs
in the Lennard-Jones chain with N=11 mobile atoms. In
Section~\ref{sec5} we consider the excitation of vibrational modes
by resonant action upon individual atoms of the chain.
Section~\ref{sec6} summarizes the main results of the paper.

\section{\label{sec2} Mathematical models describing  resonant excitation of vibrational bushes}

{We will consider two mathematical models of resonant excitation
of  bushes of nonlinear vibrational modes.}

Model 1.

We  consider monoatomic chains with free ends, which represent
atoms (\emph{terminal} atoms) of a sufficiently large mass in
comparison with that of the other atoms (\emph{internal} atoms).
The terminal atoms are acted upon by external time-periodic forces
of small amplitude.

The action of such forces leads to a gradual increase of the amplitudes of the internal atoms, starting from small ones corresponding to normal modes, up to large amplitudes in the region of significant nonlinearity of the considered system.

Model 2.

In this model we consider another way of resonant excitation of bushes of NNMs. We consider the end atoms of the monoatomic chain to be fixed, while the periodic external force acts only on one of the internal atoms of the chain. As such a chosen atom, we try successively each of the internal atoms, i.e. we take sequentially atoms with numbers 1, 2, 3, ..., $N$ in a chain consisting of $N$ mobile atoms.

Since the atom acted upon by an external periodic force pushes its
neighboring atoms, the excitation is transmitted along the entire
chain, thus leading to an excitation of some combination of normal
modes. In particular, such a combination may correspond to some
bush, i.~e. to set of NNMs, which retains its structure during the
time evolution of the system.

Let us note that the form of excitation of the chain depends on
the location of the atom on which the periodic external force
acts, but this dependence is weak.

Below, we consider resonant
excitation of different NMs, depending on the frequency of the
acting periodic force.
Let us define the concepts of permutational and inversional modes.

From any table of normal modes (see, for example Table 4), one can
note some symmetry of their patterns of atomic displacements. We  consider  the symmetry of the
displacement patterns as the symmetry of the
normal modes themselves.

We will always number normal modes in order of increasing their
frequency. For the chain with \emph{even} number of atoms, the
structure of the displacement patterns is of the form

$$[... a, b, c, d, e\,|\,e, d, c, b, a...].$$

In the center of the chain (between two central atoms) there is
the symmetry operation ($P$), which represents \emph{permutation} of the
displacements. In chains with \emph{odd} number of atoms,
permutation locates on the central atom:

$$[...a, b, c, d, c, b, a...].$$

\noindent We call modes with patterns of these type
by the term \emph{permutational modes}.

The even modes in monoatomic chains possess \emph{inversion} (operation $I$) at
the center of the chain:

$$[...a, b, c, d, 0, -d, -c, -b, -a...]$$

\noindent for the chains with the odd number of atoms and

$$[...a, b, c, d, -d, -c, -b, -a...]$$

\noindent for the chains with even number of atoms.

If inversion locates on some atom, then this atom is obviously
immobile (its displacement from the equilibrium position is zero).

Let us note that both permutational and inversional modes can possess
zero displacements on different positions that means the
presence of inversions at the corresponding places.

Note also that the aforesaid statements relates to the chains with
periodic boundary conditions. However, one can pass from this case
to the case of chains with fixed boundary conditions by formally
doubling the number of their particles (we discussed this
mathematical details in \cite{ChR2022}).

Let us consider
some examples of the above-described process of resonant
excitation of different normal modes in the monoatomic
Lennard-Jones chain.

All calculations in the present paper were performed with the homemade programs using the computational package MAPLE.

\section{\label{sec3} Excitation of vibrational modes in the chain with $N=2$ internal atoms}

Let us consider the case presented by the chain of two particles $N = 2$. Only two normal modes can exist in this system: the symmetric mode, which corresponds to the atomic displacements pattern [$a$, $a$], and the anti-symmetric mode with the pattern [$a$, $-a$]. In the symmetric normal mode, both atoms move \emph{in-phase} and in the anti-symmetric mode they move \emph{in anti-phase}.

Since the displacements patterns are defined by the eigenvectors of the force-constant matrix (matrix of second partial
derivatives of the potential energy, which are
calculated in the equilibrium state of the
considered system), they form an orthonormal set of vectors. For the chain with the Lennard-Jones interactions, the symmetric mode corresponds to the pattern [$0.7071$, $0.7071$] with circular frequency $\omega_1 = 3.7798$, while the anti-symmetric mode corresponds to the pattern [$0.7071$, $-0.7071$] with frequency $\omega_2 = 6.5467$. Hereafter, we give some numerical data with four or three decimal digits.

As was already noted in Introduction, in the framework of
computational experiments, the excitation of a given mode in the
chain with fixed ends can be carried out by setting the
appropriate initial conditions when solving nonlinear dynamical
equations by some numerical method. However, in real physical
systems, such way of excitation is impossible. Because of this
reason, we consider the model 1, which was declared in section~2. It
represents the chain with free ends, but the terminal atoms
possess a large mass, in comparison with the masses of internal
atoms, and they are acted upon by periodic external forces of
small amplitude.

In this model, the excitation of a given mode occurs due to the resonance of its frequency with the frequency of external forces.

In Ref.~\cite{ChR2022}, it was proved that an anti-symmetric mode,
being excited in a monoatomic chain with fixed ends, cannot
transfer energy (excitation) to a symmetric mode, because of which
it can persist within the framework of the considered model, for
an arbitrarily long time. On the other hand, a symmetric mode,
being excited at the initial instant, leads to the excitation of
an anti-symmetric mode already at arbitrarily small times.

It was proved in \cite{ChR2022} that an anti-symmetric mode, being
excited in the chain, cannot transfer excitation (energy) to the
symmetric mode and therefore it can exist for infinitely long
times. On the contrary, symmetric mode transfers excitation to the
anti-symmetric mode even at infinitely small times near the
initial instant $t=0$.\footnote{A similar statement holds for
chains with an arbitrary number of atoms (see \cite{ChR2022}) for
the excitation transfer between modes, one of which has a simple
permutation in the center of the chain, and the second has an
inversion in this center.}

In order to excite the normal modes with a given symmetry, we must choose the appropriate symmetry of our resonant scheme of excitation. We have two possible variants of such scheme --- the external forces act in-phase or in anti-phase.

In the first case, we can excite  symmetric modes (by appropriate choice of the frequency of external forces), while in the second case we can excite anti-symmetric modes.

In both cases, we change the frequency $\omega$ in a certain interval, where we suppose to find resonance, and for each chosen $\omega$ integrate nonlinear equations of motion for sufficiently long time from the initial instant $t=0$ for obtaining the energy $E(\omega)$ of the chain at the end of integration. The results of the below presented calculations were carried out for end time $t_f=400$, $t_f =4000$ and $t_f=20000$.

\subsection{Anti-phase excitation}

Let us begin our consideration with anti-phase
excitation. We used the following conditions for
the computer experiment. Masses of both
terminal atoms are equal 100, while those of internal atoms are equal 1.
Forces that act on the
left and right terminal atoms are $0.01 \cos(\omega t)$ and $-0.01 \cos(\omega t)$, respectively.

\begin{figure}[h!]
  \centering
  \includegraphics[width=.49\linewidth]{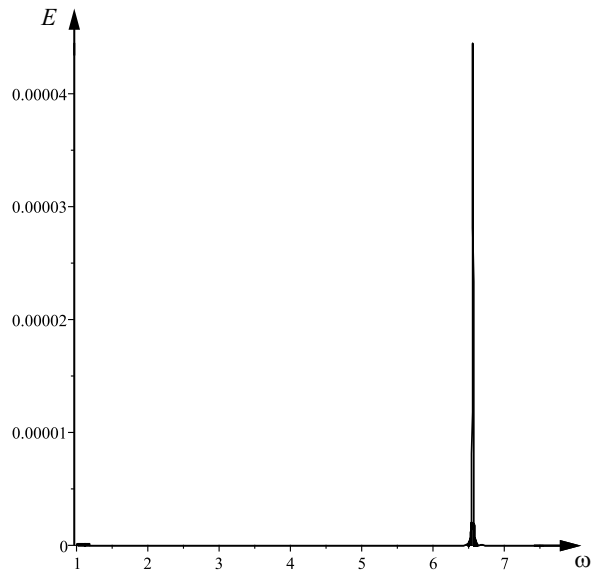}
  \caption{\label{fig1} Energy pumping into the Lennard-Jones monoatomic
chain for the frequency
$\omega_2 = 6.5467$ of the second normal mode.}
\end{figure}

Changing the frequency $\omega$ from 0 to 8 and
calculating the energy $E$ of the chain at $t=400$, we
found a sharp peak feature of the function $E(\omega) $ at
$\omega=6.5502$ depicted in Fig.~\ref{fig1}. This value corresponds
to the frequency $\omega_2=6.5467$ of the anti-symmetric
normal mode obtained in the harmonic
approximation.

\pagebreak

\begin{figure}[h!]
  \centering
  \includegraphics[width=.5\linewidth]{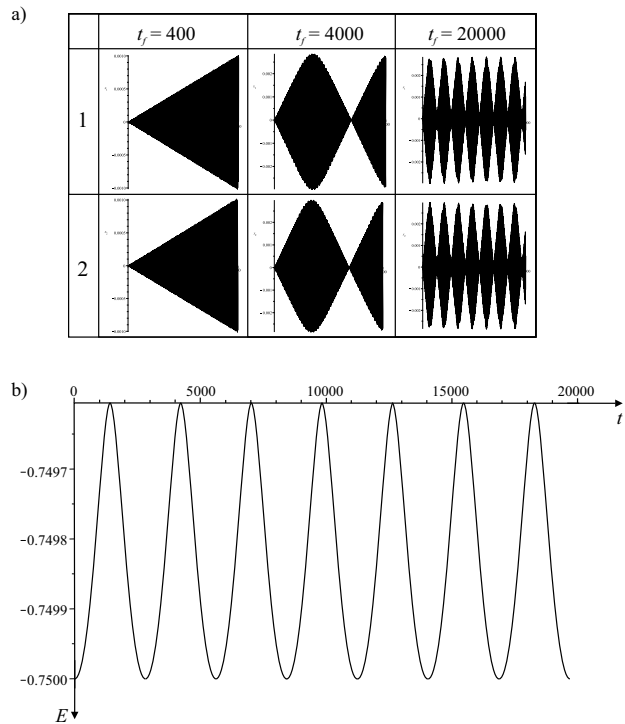}
  \caption{\label{fig4} Oscillatory behavior of the energy pumping
into the second normal mode.}
\end{figure}

In Fig.~\ref{fig4} we present plots of oscillations of both internal atoms for the chain with $N=2$.

It can be seen in this figure that the process of energy pumping
has an oscillatory nature and a period of order $T = 3000$ is
observed in time. This value of $T$ is very large in comparison
with the period of oscillations of the external forces, which is
of the order of unity.

The above phenomenon can be observed even for the shortest chain
($N=1$), which represents one atom performing anharmonic
oscillations between two terminal atoms vibrating with small
amplitudes. This problem can usually be reduced to studying of the
one-dimensional Duffing oscillator, which is acted by an external
time-periodic force. Such oscillator can exhibit very complex
behavior, up to cascades of bifurcations and transition to chaotic
dynamical regime. Among various regimes of the Duffing oscillator,
there is the small-amplitude regime of  periodic energy pumping,
completely analogous to that shown in Fig.~\ref{fig4}.

In this figure, the periodic oscillations of the energy pumped into the chain of two atoms ($N=2$) are shown. Note that the value of this energy is rather small, since the oscillation
process occurs near the harmonic approximation. Nevertheless, even in this case, the anharmonicity of the observed oscillations manifests itself quite expressively.

Let us note that in the time interval $[0, 400]$ both atoms of the two-atomic chain ($N=2$) demonstrate behavior typical for the case of exact \emph{resonance} when the frequency of the external forces coincides with the eigenfrequency of the normal mode.

The set of atomic oscillations at any fixed instant can be
expanded in the basis vectors of the force-constants matrix, which
determine the patterns of atomic displacements of the considered
modes. This is the way to construct the \emph{bush} of normal
modes excited at the given instant. In geometrical sense, such
bush represents a linear combination of a number of normal modes
whose coefficients depend on time.

\newpage

In the considered case $N=2$, any bush is a linear combination of two normal modes --- symmetric (mode 1) and anti-symmetric (mode 2).

In Table 1, we present coefficients of the
expansion of the excited dynamical object
onto above two modes for the case of in-phase acting of the external forces up to the final instants $t_f = 400$, $t_f =
400$ and $t_f =20000$.

\begin{table}[h!]
\centering
    \begin{tabular}{|l|c|c|c|}
  \hline
    & $t_f=400$ & $t_f=4000$ & $t_f=20000$ \\
    \hline
  1  & $-1.444677033167\cdot10^{-14}$ & $-1.806510019700\cdot10^{-12}$ & $-2.249376325393\cdot10^{-11}$ \\
  \hline
  2  & $0.00006576727703872$ & $0.003810472067159$ & $-0.0009959580353312$ \\
  \hline
\end{tabular}
  \caption{\label{table1} Energy pumped into the second mode at different moments.}
\end{table}

From this table, one can see that anti-symmetric mode 2 holds a
very large component of the atomic displacement pattern in
comparison with that of the symmetric mode 1. This means that we
have obtained with good computational accuracy a
\emph{one-dimensional} bush containing only the anti-symmetric
mode.

\subsection{In-phase excitation}

In this case, the external forces acting on the left and right
atoms of the chain are $0.01\cos(\omega t)$ and $0.01 \cos(\omega
t)$, respectively. The masses of the terminal atoms are equal 100,
the masses of the internal atoms are equal 1.
\begin{figure}[h!]
  \centering
  \includegraphics[width=.5\linewidth]{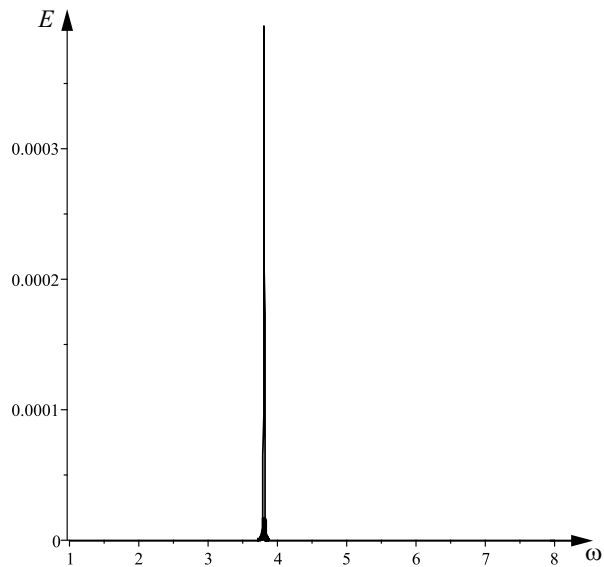}
  \caption{\label{fig2} Energy pumping into the monoatomic Lennard-Jones chain for the frequency
$\omega_1 = 3.7984$ of the first normal mode.}
\end{figure}

One can see from Fig.~\ref{fig2} that the symmetric normal mode is
excited at the frequency $\omega=3.7984$, which is close to its
eigenfrequency $\omega_1=3.7798$.

\begin{figure}[h!]
  \centering
  \includegraphics[width=.5\linewidth]{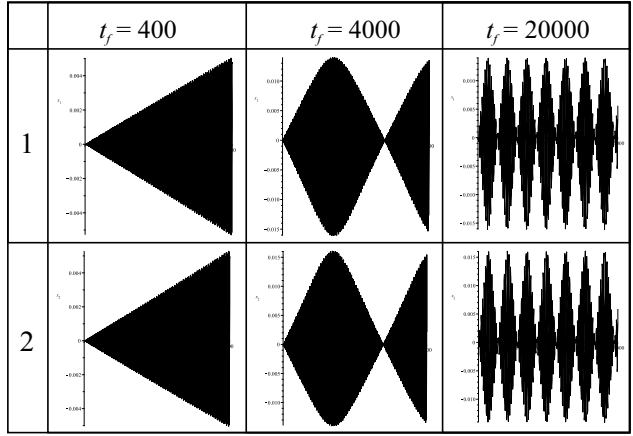}
  \caption{\label{fig4s} Oscillatory behavior of the energy pumping
into the first normal mode.}
\end{figure}

Numerical integration provides the evolution of the atomic displacements presented in Fig.~\ref{fig4s}.

The decomposition of the excited dynamical object in basis of two considered normal modes can be seen in Table 2.

\begin{table}
\centering
    \begin{tabular}{|l|c|c|c|}
  \hline
    & $t_f=400$ & $t_f=4000$ & $t_f=20000$ \\
    \hline
  1  & $0.00698593311704724$ & $0.0189079601036544$ & $0.00410249463190288$ \\
  \hline
  2  & $-0.000140942799142260$ & $-0.00107995742819299$ & $0.0000395613586320438$ \\
  \hline
\end{tabular}
  \caption{\label{table2} The energy pumped into the
Lennard-Jones chain with $N=11$ atoms for $t_f=400$, $4000$, $20000$.}
\end{table}

\newpage

The data of Table 2 show that we have obtained a two-dimensional
bush consisting of two modes --- symmetric mode, which is the root\footnote{Remember that by the term ''root mode'' we
call the mode that was excited at the initial
instant, while the modes involved into the
oscillatory process due to the interaction with
the root mode, we call ''secondary modes''.}
mode in this case, and anti-symmetric one, which is the secondary
mode.

\section{\label{sec4} The Lennard-Jones chain with N=11 internal atoms}

Let us now consider the Lennard-Jones chain with $N=11$ internal atoms.
Note that the amplitude of the secondary mode is smaller than that
of the root mode because of the week nonlinearity of the
considered chain. Let us also remember that excitation of the
symmetric mode starts to transfer to the anti-symmetric mode even
at infinitesimal times near $t=0$ (see Sec. 3.1).

\subsection{Anti-phase excitation}

We will excite the chain by acting on its terminal atoms with external periodic forces in the anti-phase regime. This means that the forces $A\cdot\cos (\omega\cdot t)$ and $-A\cdot\cos (\omega\cdot t)$ act on the left and right terminal atoms, respectively. In the computational experiments presented below, it was assumed $A=0.01$.

All displacements of the internal atoms from their equilibrium
positions and their velocities were assumed initially to be zero.

The frequency of external forces varied in the interval $[0, 8]$. For each value of the frequency $\omega$ from this interval, the nonlinear Lennard-Jones equations were integrated from zero up to a certain final instant  $t_f$, and then the energy pumped into the chain at this moment was calculated.

   \begin{figure}[h!]
  \centering
  \includegraphics[width=.5\linewidth]{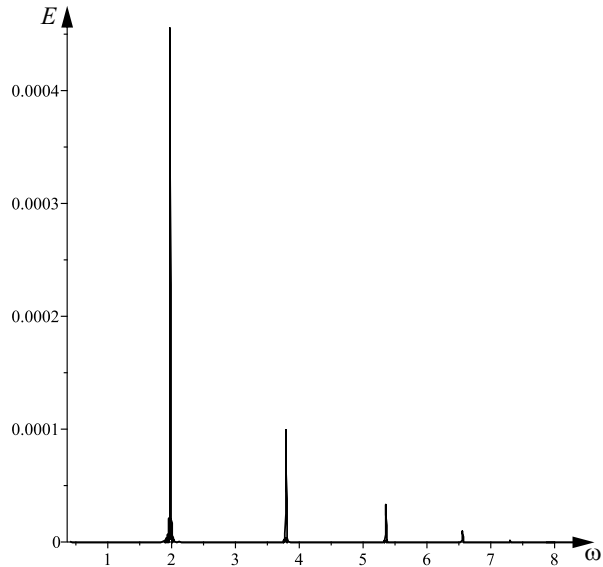}
  \caption{\label{fig10} Anti-phase energy pumping into the
Lennard-Jones chain with $N=11$ mobile atoms.}
\end{figure}

\newpage

It can be seen from  Fig.~\ref{fig10} that there is a
\emph{discrete spectrum} of external force frequencies
corresponding to the \emph{resonant pumping} of the energy into
the considered chain. These resonant frequencies are determined by
the energy peaks shown in Fig.~\ref{fig10}. With the aid of
computational experiments, we found the following resonant
frequencies for the case of anti-phase pumping: $\omega = 1.968$,
$3.784$, $5.348$, $6.548$, $7.302$. For frequencies outside the
peaks, the energy pumped into the chain is practically zero.

 From
the same Fig.~\ref{fig10}, it is also seen that the height of the
resonant peaks rapidly decreases with increase in the
corresponding frequencies.

\begin{figure}[h!]
  \centering
  \includegraphics[width=.85\linewidth]{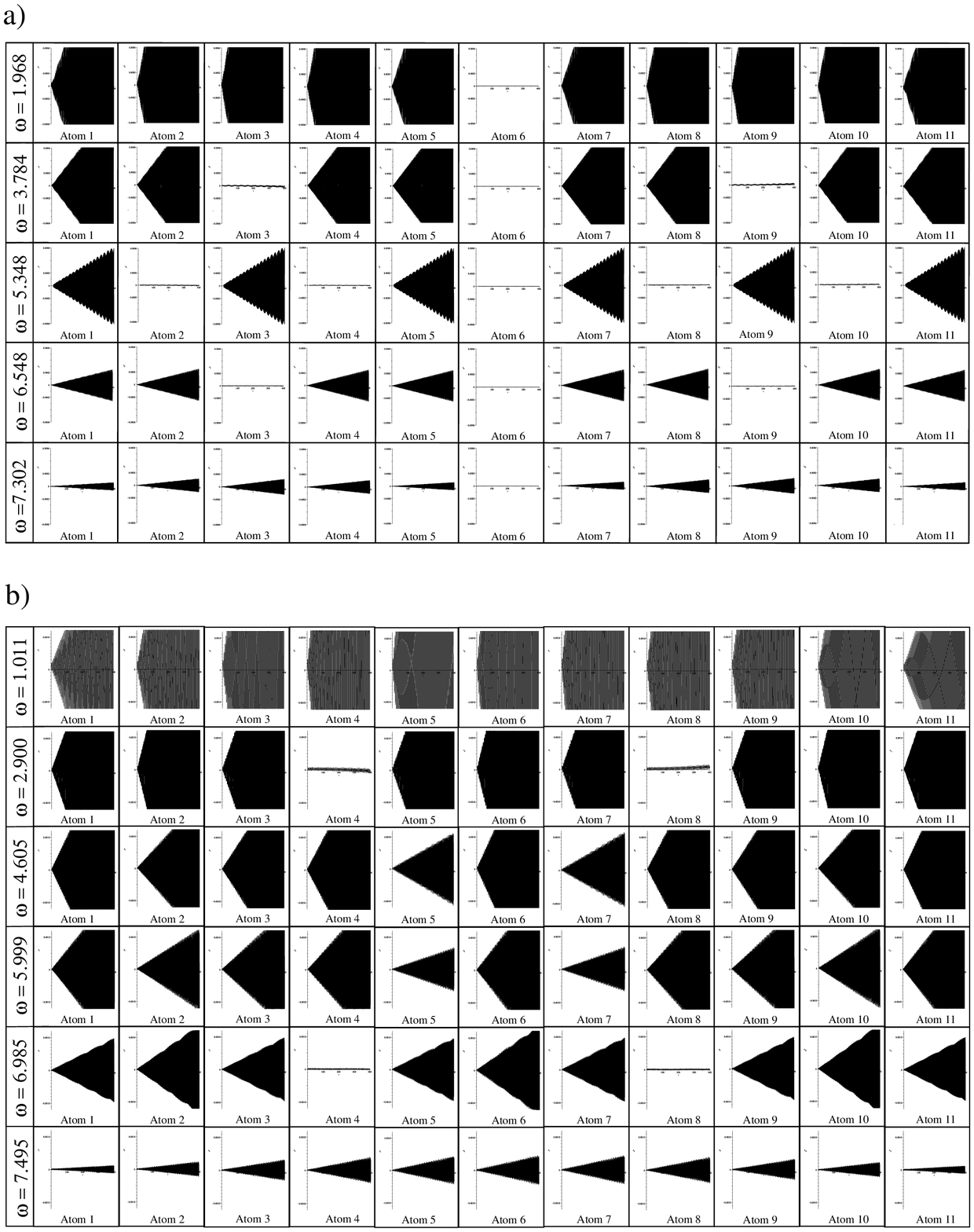}
  \caption{\label{fig11} Excitation of individual atoms by anti-phase (a) and in-phase (b) action of periodic
forces in the Lennard-Jones chain with $N=11$ mobile atoms.}
\end{figure}

In Fig.~\ref{fig11}a, we present plots of the resonant pumping
separately for each atom of the chain. It can be seen from this
figure that atoms with zero components in the displacement pattern
of a given normal mode practically do not increase their energy
during resonant pumping, in contrast to atoms with nonzero
displacements in the pattern of this mode.

We would like to note that some of the relationships between components of the displacement pattern of a given normal mode disappear as the energy is pumped into the chain.
 This is due to the involvement of some other normal modes into the vibrational process, i.e. to the formation of a certain bush of modes with  dimension greater than unity (see below).

\subsection{In-phase excitation}

In Fig.~\ref{fig11}b, we present plots of the resonant pumping separately for each atom of the chain.
To obtain this dependence we set a one-dimensional lattice in the frequency interval $[0, 8]$
with a fairly small step and for each frequency of this lattice the equations of motion were numerically integrated and the energy pumped into the chain by the time $t_f=400$ was found.

\begin{figure}[h!]
  \centering
  \includegraphics[width=.5\linewidth]{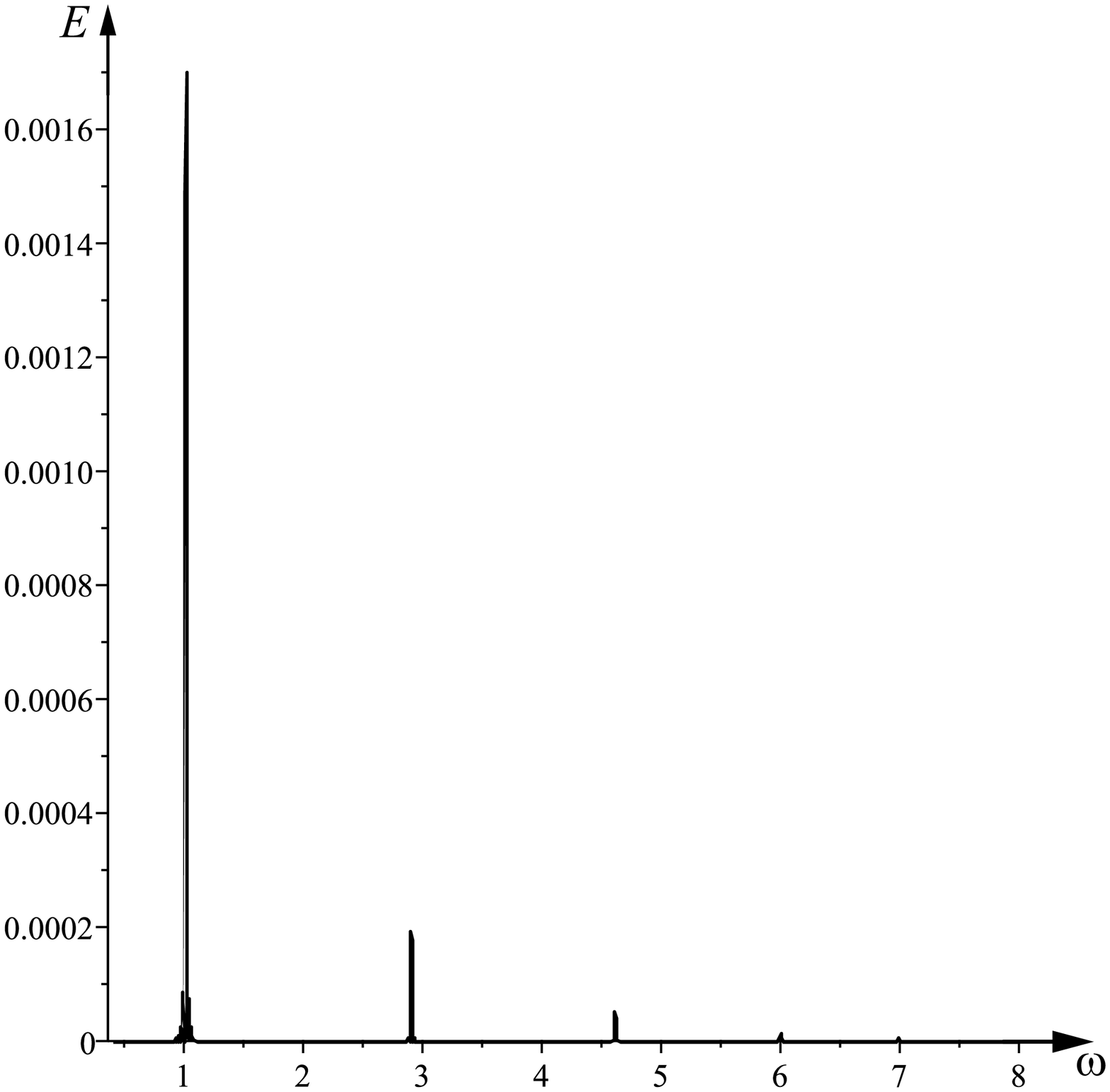}
  \caption{\label{fig10s} In-phase energy pumping into the
Lennard-Jones chain with $N=11$ mobile atoms.}
\end{figure}

The energy $E(\omega)$ pumped into the Lennard-Jones chain with $N = 11$ internal atoms at the
instant $t_f = 400$ is shown for this case in Fig.~\ref{fig10s}.
The peaks in this figure correspond to the
following set of resonant frequencies: $\omega = 1.011$, $2.900$, $4.605$, $5.999$, $6.985$, $7.495$.

It can be seen from the above figure that the value of the energy pumped into the chain strongly depends on the frequency at which the pumping was carried out, and sharply decreases with an increase in this frequency.

\newpage

In Table~\ref{table10}, we show the frequencies of all vibrational modes in the Lennard-Jones chain with $N=11$ internal atoms. The first column lists the frequencies of normal modes. Results of the computational experiments are presented in two next columns of the table. In the second column, the frequencies of vibrational modes found with the aid of anti-phase excitation scheme (in this case, even-numbered modes are excited). In the third column, the same information is presented for the case of in-phase excitation (in this case, modes with odd numbers are excited).

\begin{table}
\centering
    \begin{tabular}{|c|c|c|c|}
  \hline
N   & Frequency      & Frequency obtained & Frequency obtained   \\
     & of normal mode & by anti-phase excitation & by in-phase excitation  \\
    \hline
1&0.987 &   1.011   &   \\ \hline
2&1.957 &   &   1.968  \\ \hline
3&2.893&    2.900   &   \\ \hline
4&3.780&    &   3.784  \\ \hline
5&4.602&    4.605   &   \\ \hline
6&5.345&    &   5.348  \\ \hline
7&5.997&    5.999   &   \\ \hline
8& 6.547&   &   6.548  \\ \hline
9&6.984&    6.985&    \\ \hline
10&7.302&       & 7.302  \\ \hline
11&7.495&   7.495   &  \\ \hline
\end{tabular}
  \caption{\label{table10} Frequencies of the vibrational modes of the Lennard-Jones chain with $N=11$ internal particles .}
\end{table}

In Fig.~\ref{fig11}a and Fig.~\ref{fig11}b, we show the time
evolution of atomic displacements in the chain with $N=11$ internal
atoms from the initial instant $t=0$ up to $t_f= 400$.

Fig.~\ref{fig11} demonstrates the time evolution of all 11 atoms of the Lennard-Jones chain during pumping energy at the frequency $\omega=5.348 $ corresponding to the normal mode $6$.

This is the most symmetric mode with five immobile atoms (5 zeros in its pattern of atomic displacements).

At the initial instant $t=0$ all atoms of the chain are in
equilibrium positions. Under the action of the periodic external
forces with the frequency  rather close to $t_f=400$ that of the
normal mode 6 ($\omega=5.348$, see Table \ref{table10}), the
intense energy pumping to all atoms with even numbers begins,
while atoms with odd numbers display an energy pumping rate that
is three orders of magnitude lower.

\newpage

As another example, Fig.~\ref{fig11} shows the time evolution of
 energy pumping to the atoms of the chain at the frequency
corresponding to the normal mode 4 ($\omega=3.784$). The pattern
of atomic displacements of this mode has three immobile atoms with
numbers 3, 6 and 9. Again we see that atoms, which were immobile
in the chain with fixed ends, possess in the case of exact
resonance three orders of magnitude lower energy pumping rate
compared to that corresponding to nonzero components of the atomic
displacement pattern.

Thus, we see that the correspondence of the excited vibrational mode to a certain normal mode can be established both by the proximity of its frequency to that of the normal mode and by the rate of energy pumping from atoms with zero and nonzero components of the atomic displacement pattern of this normal mode.

The atomic motion during the resonance pumping up to $t_f=400$ is quite close to that in the harmonic approximation because of the smallness of atomic displacements.

\section{\label{sec5} Excitation of nonlinear vibrational modes in the case of the resonant action on individual atoms of the chain}

In the previous part
of the work, we have studied the model of monoatomic
chain whose end atoms
possess significantly larger
mass compared to that of the
inner atoms of the chain.
Periodic external forces with
small amplitude and with
 resonant frequency
acted on the end atoms of the chain. Two different schemes of
energy pumping into the chain were studied. They are anti-phase
and in-phase pumping.

In the first case,
all odd normal modes
and bushes generated
by these root modes
can be excited
by resonant action,
while in the second
case, all even modes
can be excited.

Below, we consider another way of resonant excitation of bushes of
NNMs represented by the model 2 We consider the end atoms of the
monoatomic chain to be fixed, while the periodic external force
acts only on one of the internal atoms of the chain. For the
chosen atom, we try successively each of the internal atoms, i.e.
we take sequentially atoms numbers 1, 2, 3, ..., $N$ in a chain
consisting of $N$ mobile atoms.

Since the atom acted upon by an external periodic force, pushes
its neighboring atoms, the excitation is transmitted along the
entire chain, thus leading to the excitation of some combination
of normal modes. In particular, such a combination may correspond
to some bush, i.~e. to set of NNMs, which retains its structure
during the time evolution of the system.

Let us note that the form of excitation of the chain depends on
the location of the atom on which the periodic external force
acts, but this dependence is weak.

Let us consider some examples of the above-described process of resonant excitation of different normal modes in the monoatomic Lennard-Jones chain.

\tabcolsep=2pt

\begin{table}
\centering
\begin{tabular}{|c|c|c|c|c|c|c|c|c|c|c|}
  \hline
 1 &    2 &    3 &    4 &    5 &    6 &    7 &    8 &  9   &  10   &  11 \\ \hline
$\omega = 1.011$ &  $\omega = 1.968$ &  $\omega = 2.900$ &  $\omega = 3.784$ &  $\omega = 4.605$ &  $\omega = 5.348$ &  $\omega = 5.999$ &   $\omega = 6.548     $ &    $\omega =  6.985$ & $\omega = 7.302$ &   $\omega = 7.495$ \\ \hline
0.106   &0.204  &0.289  & 0.354  & 0.394    &0.408 & 0.394  &  0.354  & 0.289   &0.204  &0.106 \\ \hline
0.204   &0.354  &0.408  & 0.354  & 0.204    &0      &-0.204 & -0.354  & -0.408  &-0.354 &-0.204 \\ \hline
0.289   &0.408  &0.289  & 0      & -0.289&  -0.408& -0.289  & 0       &  0.289  &0.408  &0.288 \\ \hline
0.354   &0.354  &0      & -0.354 & -0.354&  0    &  0.354   & 0.354   & 0       &-0.354 &-0.354 \\ \hline
0.394   &0.204  &-0.289&  -0.354&  0.107&   0.408&  0.106   & -0.354  & -0.289  &0.204  &0.394 \\ \hline
0.408   &0.     &-0.408&  0    &   0.408&   0     &  -0.408 &   0     &  0.408  &0      &-0.408 \\ \hline
0.394   &-0.204 &-0.289& 0.354  &  0.106&   -0.408& 0.106   & 0.354   & -0.289  &-0.204 &   0.394 \\ \hline
0.354   &-0.354 &0    &  0.354  &  -0.354&  0     &  0.354  & -0.354  & 0       &0.354  &-0.354 \\ \hline
0.289   &-0.408 &0.289&  0      &  -0.289&  0.408&  -0.289  & 0       & 0.289   &-0.408 &0.289 \\ \hline
0.204   &-0.354 &0.408&  -0.354&   0.204&   0     &  -0.204 & 0.354   & -0.408  &0.354  &-0.204 \\ \hline
0.106   &-0.204 &0.289&  -0.354&   0.394&   -0.408& 0.394   & -0.354  & 0.289   &-0.204 &0.106 \\ \hline
\end{tabular}
  \caption{\label{table40} Normal modes for monoatomic chain with
$N=11$ mobile atoms for the case of Lennard-
Jones interactions.}
\end{table}

\tabcolsep=5pt

\subsection{Mode 1}

\begin{figure}[h!]
  \centering
  \includegraphics[width=.64\linewidth]{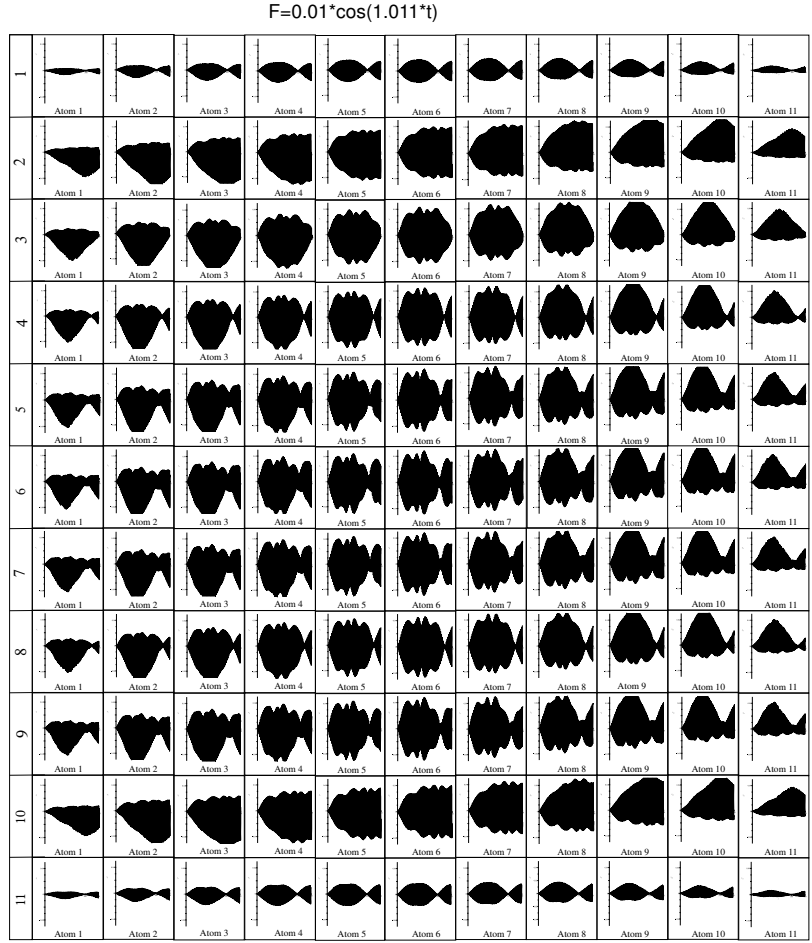}
  \caption{\label{figw26} The resonant excitation of atomic vibrations in the monoatomic Lennard-Jones
chain with $N=11$ mobile atoms under the action of the periodic external force with frequency of the
first normal mode.}
\end{figure}

Fig.~\ref{figw26} illustrates resonant pumping of the energy into the chain with $N=11$ mobile atoms as a result of the action on the chain by periodic external force at the frequency $\omega=1.0111$ of the first mode when this force is applied to \emph{any} atom of the chain. This pumping is expressed in a characteristic increase in the amplitude of oscillations of all atoms, resembling little triangles.

Each line $j=1$, $2$, $3$, $4$, $5$ in the Fig.~\ref{figw26}
corresponds to the case of an external force acting on the atom
with number $j$. It is clearly seen from this figure that the form
of atomic vibrations for any $j$ remains qualitatively the same.

\newpage

\subsection{Mode 2}

It was already said that all odd modes are permutational, since a simple permutation sits in the center of their patterns, while all even modes are inversional, since their patterns possess inversion in the center of the chain. Thus, mode 2 is the first inversional mode in the considered chain.

It can be seen from the displacement patterns (see Table~\ref{table40}) that this normal mode has one immobile atom. This is the central atom of the chain with number 6.

\begin{figure}[h!]
  \centering
  \includegraphics[width=.6\linewidth]{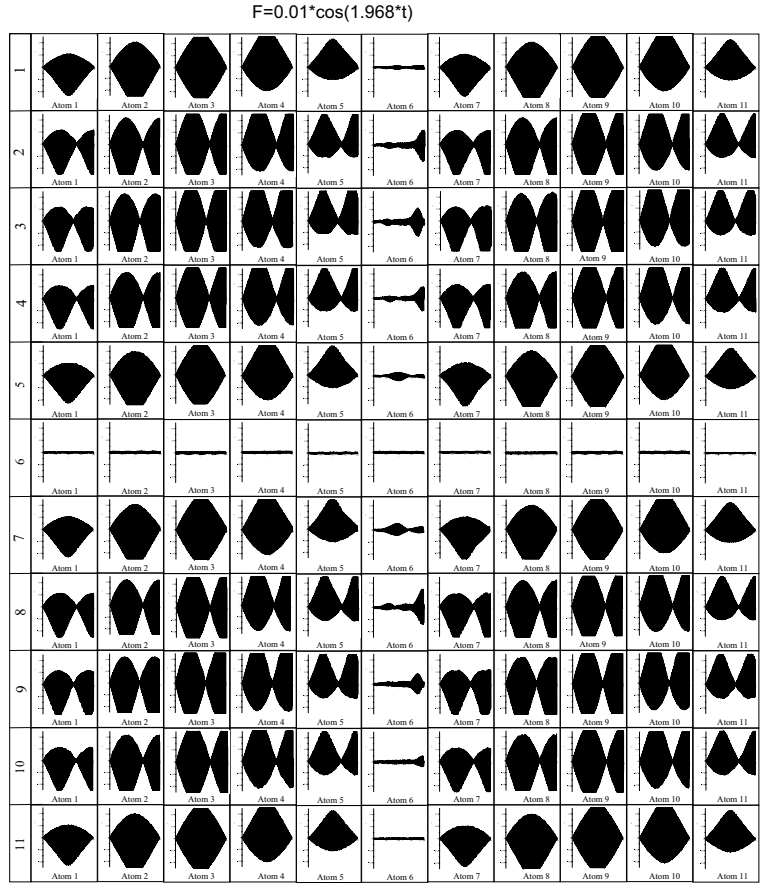}
  \caption{\label{figw52} The resonant excitation of atomic vibrations in the monoatomic Lennard-Jones
chain with $N=11$ mobile atoms under the action of the periodic external force with frequency of the
second normal mode.}
\end{figure}

Let's see how this shows up in Fig.~\ref{figw52}.

In this case, the frequency of the external force coincides with the frequency of mode 2 and we start integration with the pattern of this normal mode. As a result, resonant pumping of energy into the chain occurs, which manifests itself in the appearance of a ''triangular'' dependence of amplitudes of atomic oscillations on time during the evolution of the system. On the other hand, atom 6, which is immobile in the pattern of mode 2, performs only small oscillations around its equilibrium position.

\newpage

\subsection{Mode 3}

Being odd, this mode is permutational. However, there are two immobile atoms in its pattern (see table~\ref{table40}), which correspond to zeros of this pattern. These are atoms 4 and 8.

\begin{figure}[h!]
  \centering
  \includegraphics[width=.6\linewidth]{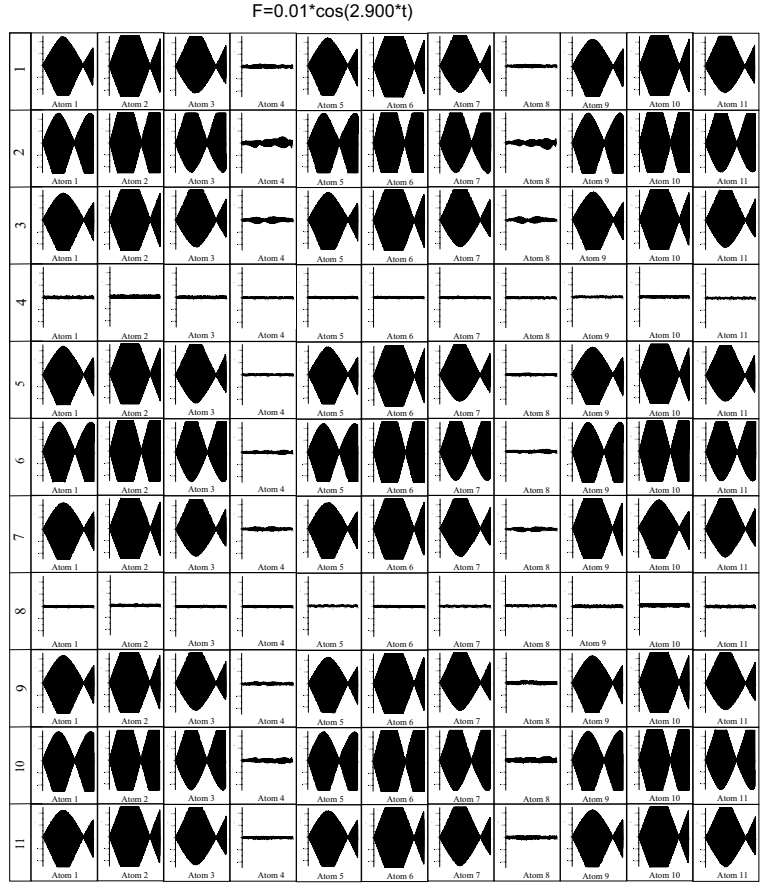}
  \caption{\label{figw76} The resonant excitation of atomic vibrations in the monoatomic Lennard-Jones
chain with $ N=11$ mobile atoms under the action of the periodic external force with frequency of the
third normal mode.}
\end{figure}

The presence of such atoms leads to characteristic features of the
resonant energy pumping into the chain. It can be seen from
Fig.~\ref{figw76} that atoms 4 and 8 themselves oscillate only
with a small amplitude proportional to the amplitude of the
external periodic force, while all other atoms demonstrate a
resonant increase of oscillation amplitudes during the time
evolution of the system. Hereafter, atoms determining the picture
of the resonant pumping of the energy into the chain, which
correspond to zeros of the pattern of considered mode, will be
called the ''\emph{key atoms}''.

\newpage

\subsection{Mode 4}

\begin{figure}[h!]
  \centering
  \includegraphics[width=.6\linewidth]{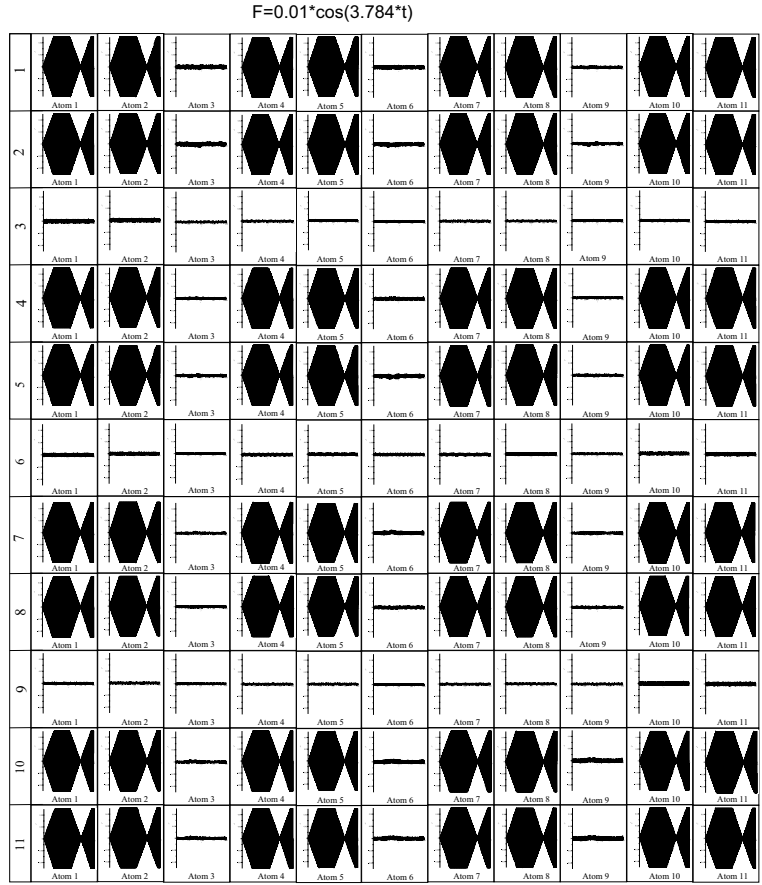}
  \caption{\label{figw1} The resonant excitation of atomic vibrations in the monoatomic Lennard-Jones
chain with $N=11$ mobile atoms under the action of the periodic external force with frequency of the
fourth normal mode.}
\end{figure}

Key atoms for the resonant pumping of energy into the chain at the frequency $\omega= 1.000$ of mode 4 are atoms $3$, $6$, $9$. These three atoms determine the rows and columns of Fig.~\ref{figw1} in which small vibrations of these atoms are visible against the background of the resonant increasing of the vibrational amplitudes of all other atoms.

\newpage

\subsection{Mode 5}

Being odd, this mode is permutational and its
displacements pattern does not contain any zeros. The consequence of this is the resonant
excitation of all atoms of the chain.

\begin{figure}[h!]
  \centering
  \includegraphics[width=.6\linewidth]{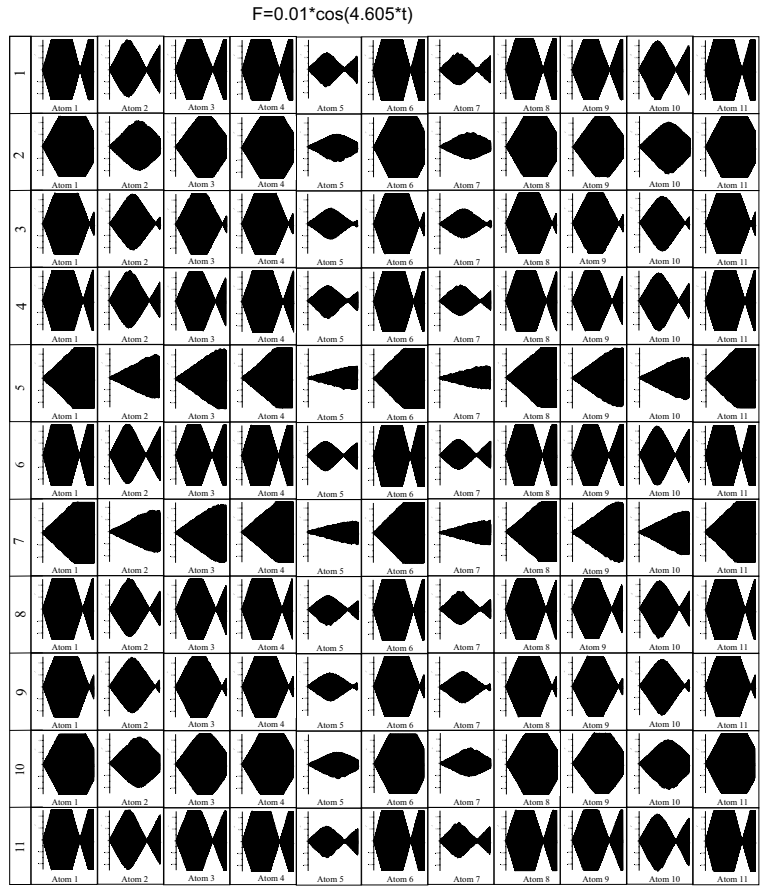}
  \caption{\label{figw121} The resonant excitation of atomic vibrations in the monoatomic Lennard-Jones
chain with $N=11$ mobile atoms under the action of the periodic external force with frequency of the
fifth normal mode.}
\end{figure}

\newpage

\clearpage

\section{\label{sec6} Summary}

In the present paper, we discuss
the problem of excitation of
bushes of nonlinear vibrational
modes using resonant action on
the atoms of monoatomic chains.
The importance of this problem
is connected with the importance
of the very concept of bushes of
nonlinear normal modes (NNM) in
arbitrary systems with discrete
symmetry at nano-, micro-, meso-
and macro-levels (in particular,
in molecular and crystal
structures with arbitrary type
of interatomic interactions).

Any bush is a certain set
of nonlinear normal modes (NNMs)
whose structure is preserved in
time. In the geometric sense, it
represents some invariant
manifold in the phase space,
expanded over the complete set
of all nonlinear normal modes of
the given system with discrete
symmetry under the assumption of
Lyapunov`s stability of motion.
This manifold corresponds to a
certain subgroup of the symmetry
group of the system in
equilibrium state(or the group
of its Hamiltonian), due to
which it does not depend on the
type and magnitude of the
interatomic interactions in the
system up to the loss of
stability of motion due to the
internal parametric resonance.
The latter property makes it
possible to consider atomic
oscillations of large
amplitudes, which cannot be done within the framework of any
perturbation theory.

In this work, we study the
excitation of bushes of NNMs in
one-dimensional monoatomic
chains due to the resonant
action on the atoms of the chain
in the framework of two
mathematical models (models 1
and 2). In the first case, a
periodic external force acts, at
the frequency of some normal
mode, on the terminal atoms of
the chain, while in the second
case it acts on individual
internal atoms. In both cases,
these methods make it possible
to excite all bushes of
nonlinear vibrational modes in
one-dimensional monoatomic
chains.
It is important that the
second method can be used to
excite bushes of NNMs in the
systems with discrete symmetry
of arbitrarily complex atomic
structure.
We will discuss this problem in detail
elsewhere.

%\section*{Acknowledgments}

\section*{ }

\bibliography{R_e2022_arxiv}

\begin{thebibliography}{10}
\expandafter\ifx\csname url\endcsname\relax
  \def\url#1{\texttt{#1}}\fi
\expandafter\ifx\csname urlprefix\endcsname\relax\def\urlprefix{URL }\fi
\expandafter\ifx\csname href\endcsname\relax
  \def\href#1#2{#2} \def\path#1{#1}\fi

\bibitem{FPU1955}
E.~Fermi, J.~Pasta, S.~M. Ulam, Studies of nonlinear problems.~{I}, Technical
  Report LA-1940, LANL, also in {\em Enrico Fermi: Collected Papers, Vol.~2},
  edited by E.~Amaldi, H.~L.~Anderson, E.~Persico, E.~Segr{\'e}, and
  A.~Wattenberg (University of Chicago Press, Chicago, 1965) pp. 978--988. (May
  1955).

\bibitem{ZabuskyKruskal1965}
N.~J. {Zabusky}, M.~D. {Kruskal}, {Interaction of ``solitons'' in a
  collisionless plasma and the recurrence of initial states}, Phys. Rev. Lett.
  15~(6) (1965) 240--243.
\newblock \href {http://dx.doi.org/10.1103/PhysRevLett.15.240}
  {\path{doi:10.1103/PhysRevLett.15.240}}.

\bibitem{Zabusky1981}
N.~J. {Zabusky}, {Computational synergetics and mathematical innovation}, J.
  Comput. Phys. 43~(2) (1981) 195--249.
\newblock \href {http://dx.doi.org/10.1016/0021-9991(81)90120-0}
  {\path{doi:10.1016/0021-9991(81)90120-0}}.

\bibitem{Chaos2005}
T.~{Dauxois}, R.~{Khomeriki}, F.~{Piazza}, S.~{Ruffo}, {The anti-FPU problem},
  Chaos 15 (2005) 015110.
\newblock \href {http://dx.doi.org/10.1063/1.1854273}
  {\path{doi:10.1063/1.1854273}}.

\bibitem{Ruffo1997}
P.~{Poggi}, S.~{Ruffo}, {Exact solutions in the FPU oscillator chain}, Physica
  D 103~(1-4) (1997) 251--272.
\newblock \href {http://dx.doi.org/10.1016/S0167-2789(96)00262-X}
  {\path{doi:10.1016/S0167-2789(96)00262-X}}.

\bibitem{art23}
S.~{Shinohara}, {Low-dimensional solutions in the quartic Fermi-Pasta-Ulam
  system}, J. Phys. Soc. Jpn. 71~(8) (2002) 1802.
\newblock \href {http://dx.doi.org/10.1143/JPSJ.71.1802}
  {\path{doi:10.1143/JPSJ.71.1802}}.

\bibitem{art24}
S.~{Shinohara}, {Low-dimensional subsystems in anharmonic lattices}, Prog.
  Theor. Phys. Supplement 150 (2003) 423--434.
\newblock \href {http://dx.doi.org/10.1143/PTPS.150.423}
  {\path{doi:10.1143/PTPS.150.423}}.

\bibitem{art25}
K.~{Yoshimura}, {Modulational instability of zone boundary mode in nonlinear
  lattices: Rigorous results}, Phys. Rev. E 70~(1) (2004) 016611.
\newblock \href {http://dx.doi.org/10.1103/PhysRevE.70.016611}
  {\path{doi:10.1103/PhysRevE.70.016611}}.

\bibitem{art26}
A.~{Cafarella}, M.~{Leo}, R.~A. {Leo}, {Numerical analysis of the one-mode
  solutions in the Fermi-Pasta-Ulam system}, Phys. Rev. E 69~(4) (2004) 046604.
\newblock \href {http://dx.doi.org/10.1103/PhysRevE.69.046604}
  {\path{doi:10.1103/PhysRevE.69.046604}}.

\bibitem{art27}
B.~{Rink}, {Symmetry and resonance in periodic FPU chains}, Commun. Math. Phys.
  218~(3) (2001) 665--685.
\newblock \href {http://dx.doi.org/10.1007/s002200100428}
  {\path{doi:10.1007/s002200100428}}.

\bibitem{art28}
B.~{Rink}, {Symmetric invariant manifolds in the Fermi-Pasta-Ulam lattice},
  Physica D 175~(1-2) (2003) 31--42.
\newblock \href {http://dx.doi.org/10.1016/S0167-2789(02)00694-2}
  {\path{doi:10.1016/S0167-2789(02)00694-2}}.

\bibitem{pss1989}
G.~{Chechin}, T.~{Ivanova}, V.~{Sakhnenko}, {Complete order parameter
  condensate of low- symmetry phases upon structural phase transitions},
  Physica Status Solidi (B) 152~(2) (1989) 431--446.
\newblock \href {http://dx.doi.org/10.1002/pssb.2221520205}
  {\path{doi:10.1002/pssb.2221520205}}.

\bibitem{art10}
G.~M. {Chechin}, V.~P. {Sakhnenko}, {Interactions between normal modes in
  nonlinear dynamical systems with discrete symmetry. Exact results}, Physica D
  117~(1-4) (1998) 43--76.
\newblock \href {http://dx.doi.org/10.1016/S0167-2789(98)80012-2}
  {\path{doi:10.1016/S0167-2789(98)80012-2}}.

\bibitem{art9}
V.~P. {Sakhnenko}, G.~M. {Chechin}, {Symmetrical selection rules in nonlinear
  dynamics of atomic systems}, Physics Doklady 38~(5) (1993) 219--221.

\bibitem{ChR2022}
G.~M. {Chechin}, D.~S. {Ryabov}, {Exact solutions of nonlinear dynamical
  equations for large-amplitude atomic vibrations in arbitrary monoatomic
  chains with fixed ends. }\href
  {http://dx.doi.org/https://arxiv.org/abs/2109.09101}
  {\path{doi:https://arxiv.org/abs/2109.09101}}.

\bibitem{art14}
G.~M. {Chechin}, A.~V. {Gnezdilov}, M.~Y. {Zekhtser}, {Existence and stability
  of bushes of vibrational modes for octahedral mechanical systems with
  Lennard-Jones potential}, Int. J. Non-Linear Mechanics 38~(10) (2003)
  1451--1472.
\newblock \href {http://dx.doi.org/10.1016/S0020-7462(02)00081-1}
  {\path{doi:10.1016/S0020-7462(02)00081-1}}.

\bibitem{art15}
G.~{Chechin}, D.~{Ryabov}, S.~{Shcherbinin}, {Nonlinear normal mode
  interactions in the SF$_{6}$ molecule studied with the aid of density
  functional theory}, Phys. Rev. E 92~(1) (2015) 012907.
\newblock \href {http://dx.doi.org/10.1103/PhysRevE.92.012907}
  {\path{doi:10.1103/PhysRevE.92.012907}}.

\bibitem{art17}
G.~M. {Chechin}, D.~S. {Ryabov}, S.~A. {Shcherbinin}, {Large-amplitude in-plane
  atomic vibrations in strained graphene monolayer: Bushes of nonlinear normal
  modes}, Letters on Materials 7~(4) (2017) 367--372.
\newblock \href {http://dx.doi.org/10.22226/2410-3535-2017-4-367-372}
  {\path{doi:10.22226/2410-3535-2017-4-367-372}}.

\bibitem{art18}
G.~{Chechin}, D.~{Ryabov}, S.~{Shcherbinin}, {Large-amplitude periodic atomic
  vibrations in diamond}, Journal of Micromechanics and Molecular Physics
  03~(01n02) (2018) 1850002.
\newblock \href {http://dx.doi.org/10.1142/S2424913018500029}
  {\path{doi:10.1142/S2424913018500029}}.

\bibitem{art34}
G.~M. {Chechin}, S.~V. {Dmitriev}, I.~P. {Lobzenko}, D.~S. {Ryabov},
  {Properties of discrete breathers in graphane from ab initio simulations},
  Phys. Rev. B 90~(4) (2014) 045432.
\newblock \href {http://dx.doi.org/10.1103/PhysRevB.90.045432}
  {\path{doi:10.1103/PhysRevB.90.045432}}.

\bibitem{art51}
G.~M. {Chechin}, K.~G. {Zhukov}, {Stability analysis of dynamical regimes in
  nonlinear systems with discrete symmetries}, Phys. Rev. E 73~(3) (2006)
  036216.
\newblock \href {http://dx.doi.org/10.1103/PhysRevE.73.036216}
  {\path{doi:10.1103/PhysRevE.73.036216}}.

\bibitem{art16}
G.~M. {Chechin}, D.~A. {Sizintsev}, O.~A. {Usoltsev}, {Nonlinear atomic
  vibrations and structural phase transitions in strained carbon chains},
  Computational Materials Science 138 (2017) 353--367.
\newblock \href {http://dx.doi.org/10.1016/j.commatsci.2017.07.004}
  {\path{doi:10.1016/j.commatsci.2017.07.004}}.

\bibitem{art102}
G.~M. {Chechin}, V.~S. {Lapina}, {Universal bi-structures of strained
  monoatomic chains}, Phys. Rev. E 122 (2020) 114155.
\newblock \href {http://dx.doi.org/10.1016/j.physe.2020.114155}
  {\path{doi:10.1016/j.physe.2020.114155}}.

\bibitem{art103}
G.~M. {Chechin}, V.~S. {Lapina}, {Discrete breathers of new type in monoatomic
  chains}, Letters on Materials 8 (2018) 458--462.
\newblock \href {http://dx.doi.org/10.22226/2410-3535-2018-4-458-462}
  {\path{doi:10.22226/2410-3535-2018-4-458-462}}.

\bibitem{art104}
G.~M. {Chechin}, V.~S. {Lapina}, {Static structures of strained carbon chains:
  DFT-modeling vs classical modeling of the chain with Lennard-Jones
  potential}, Letters on Materials 9 (2019) 151--156.
\newblock \href {http://dx.doi.org/10.22226/2410-3535-2019-2-151-156}
  {\path{doi:10.22226/2410-3535-2019-2-151-156}}.

\end{thebibliography}

\end{document}